\documentclass[12pt]{article}
\usepackage[totalwidth=460truept,totalheight=628truept]{geometry}
\usepackage{latexsym,graphicx,amsfonts,amssymb,amsmath}
\usepackage{xcolor}
\usepackage[hypertexnames=false,hidelinks]{hyperref}

\def\theequation{\arabic{section}.\arabic{equation}}

\renewcommand{\theequation}{\thesection.\arabic{equation}}
\global\arraycolsep=1truept
\renewcommand{\theequation}{\arabic{section}.\arabic{equation}}

\begin{document}

\title{\huge\textbf{Causality Bounds in Quadratic Inflation from Purely Virtual Particles}}
\date{}

\author{Alessandro Dondarini}
\maketitle

\begin{center}

\textit{Dipartimento di Fisica \textquotedblleft Enrico Fermi", Universit%
\`{a} di Pisa}

\textit{Largo B. Pontecorvo 3, 56127 Pisa, Italy}

\textit{and INFN, Sezione di Pisa,}

\textit{Largo B. Pontecorvo 3, 56127 Pisa, Italy}

alessandro.dondarini@phd.unipi.it

\vskip1truecm

\textbf{Abstract}
\end{center}

The  \lq\lq$\phi^2$" slow roll inflation combined with General Relativity is largely excluded by Planck data. In this paper, we consider the same potential combined with the $R+C^2$ gravity of purely virtual particles (or \textit{fakeons}), where the would-be ghost introduced by the Weyl tensor term, $C^2$, is quantized with the fakeon prescription. We compute the tensor power spectrum  in the full theory by means of the Cosmic Renormalization Group formalism and critically examine its physical meaning. In particular, we show that it is not possible to retrieve the power spectrum of the fakeon free-theory by considering the decoupling limit of the purely virtual particles. We provide a physical explanation in terms of the causal structure of the theory to infer that a model of quadratic inflation from purely virtual particles is also discarded from a phenomenological point of view.

\vfill\eject

\section{Introduction}
\label{intro}\setcounter{equation}{0}
The current high-energy physics scenario is characterized by a mostly consistent theoretical description, whose direct implications have been tested to a high degree of accuracy. The Standard Model of particle physics is one of the greatest examples of successful theory confirmed by a large amount of experimental tests. Its powerful predictivity heavily relies on three basic principles: unitarity, locality and renormalizability. Despite its success, the Standard Model can actually furnish a satisfactory explanation of only three of the four known fundamental interactions: the electromagnetic, the weak and the strong force. Gravity, whose effects range from the smallest to the largest scales,  has not been included up to now. As a matter of fact, building a theory of gravity that respects the aforementioned principles is a hard task: General Relativity is a non-renormalizable quantum field theory \cite{thooft, sagnotti}, thus lacking of predictivity for all the energy scales\footnote{Nevertheless, General Relativity can be considered as an \textit{effective field theory} with predictions at low energy \cite{donoghue}.}. On the other hand, the simplest renormalizable extension of General Relativity \cite{stelle} is not unitary: the higher-derivative term build with the Weyl tensor, $C^2,$ propagates  a spin-2 \textit{ghost} particle when the theory is quantized with the usual $i\epsilon$ prescription.

 Recently, a new theory of quantum gravity has been formulated \cite{LWgrav, UVQG, Absograv, wheelerons, classicization, LWFormulation, LWUnitarity}. This theory heavily relies on the concept of \textit{purely virtual quantum}, a particle that can only propagate inside the  Feynman diagrams but  cannot appear as an asymptotic state. The theory circumvents the problems of Stelle's theory \cite{stelle} by means of a new approach based on the combination of a prescription and a projection, which allows us to quantize the ghost as a \textit{purely virtual particle} (or \textit{fakeon}), and then project to the physical subspace by removing it from the physical spectrum. This way, we switch to a different, ghost-free theory, and gain unitarity. In the end, the theory, which is described by the renormalizable $R+R^2+C^2$ lagrangian, propagates three degrees of freedom: a massless spin-2 particle (the graviton), a massive spin-2 field (the fakeon) and a scalar field (interpreted in the cosmological context as the inflaton, see below). Other approaches to higher-derivative quantum gravity retain the ghost particle  and give predictions in cosmology (see e.g. \cite{strumia, salvio, salvioRev}), we critically compare them with the results of section \ref{tensorQG}.
 
 The introduction of purely virtual particles modifies the usual notion of causality, leading to \textit{microcausality violations}. Such causality violations occur in a time interval $\Delta t$ and are related to the fakeon mass $m_{\chi}$, $\Delta t\sim 1/m_{\chi}$ \cite{classicization}. As a consequence, the causality violations are suppressed since the fakeon mass is expected to be very large, O$(10^{13})\,\text{GeV}$ from cosmological constraints \cite{ABP}. The goal of this paper is to show that there are exceptions. Indeed, we show that for particular inflationary dynamics the causality violation introduced by purely virtual particles propagates to large time scales. 
 
 In order to assess the phenomenological impact of said particles, we focus on in inflationary cosmology, where a number of predictions have already been derived \cite{ABP, CMBrunning, HighOrder, Last}. In this context, the scalar degree of freedom is interpreted as the inflaton field, which undergoes a phase of slow roll dynamics \cite{Guth}. The underlying inflationary potential is constrained by renormalizability and turns out to be the \textit{Starobinsky potential} \cite{starobinsky, vilenkin}. In this case, the predictivity of the theory is enforced by a consistency condition on the fakeon projection. Such condition imposes a bound on the fakeon and inflaton masses and keeps the causality violation under control by demanding that the fakeon Green function has no \textit{tachyonic poles} \cite{ABP}. The computation of the testable quantities, such as the \textit{tensor} and \textit{scalar power spectra}, are derived by means of a new formalism, the \textit{Cosmic Renormalization Group} (RG) \cite{CMBrunning}, whose features closely resemble the usual RG structure of high-energy physics. This formalism allows to compute high-order corrections to the aforementioned quantities in a consistent way and substantially improves the computational methods developed in fakeon-free theories \cite{run1, run2}. In particular, the cosmic RG flow sets a perturbative expansion for the power spectra in terms of a \textit{slow roll coupling}, which we refer as the \textit{de Sitter perturbative expansion}. As the usual QFT perturbative expansion, the de Sitter perturbative expansion is organized in terms of leading (LL), next-to-next-to-leading (NLL), next-to-leading-log (NNLL)... contributions coming from the running coupling.

 In addition to the potential derived from the theory, we can also investigate the phenomenological consequences of purely virtual particles in other inflationary scenarios. 
 
 The Starobinsky potential is obtained from the $R+R^2+C^2$ theory by means of a Weyl transformation that casts the action in the $R+C^2+V_{\rm Star}(\phi)$ form \cite{ABP}, $\phi$ and $V_{\rm Star}(\phi)$ being the inflaton field and the Starobinsky potential respectively \cite{starobinsky, vilenkin}. As an interesting possibility, we can also investigate the phenomenological consequences of purely virtual particles in other inflationary scenarios by replacing the Starobinsky potential with a  generic potential $V(\phi)$ and derive the corresponding predictions by means of the Cosmic RG formalism\footnote{See \cite{Last} for a systematic study.}. In this way, we obtain a theory that is completely different from the original $R+R^2+C^2$ theory both from the formal and phenomenological perspective. In this paper, we consider the scenario of \textit{quadratic inflation}, where $V(\phi)\propto \phi^2$. First, we derive the predictions in the limit of infinitely heavy fakeon, where the theory reduces to General Relativity combined with the inflaton action. In this case, we make a comparison with Planck data \cite{Planck18} and recover the well known conclusion: quadratic inflation is \textit{excluded} by the bound on the \textit{tensor to scalar ratio}. Then, we derive the physical predictions of the full theory and show that the consistency condition for the fakeon projection is violated within the validity of the de Sitter perturbative expansion, so that the theory is no longer causal on large time scales. This fact is reflected in the structure of the coefficients in the power spectra, which exhibits unusual physical features. In particular, we show that the decoupling limit of the fakeon does not retrieve the results of the fakeon-free theory, General Relativity, but leads to divergent power spectra as $m_{\chi}\to\infty$. As we show in the last section, the natural explanation of this fact is that we are unable to retrieve a causal theory from a strongly \textit{acausal} theory. As a side note, we point out that the $\lq\lq\phi^2"$ potential from purely virtual particles falls in the class of potentials discussed in \cite{Last}. However, Ref. \cite{Last} does not provide the explicit form of the coefficients appearing in the tensor power spectrum since it aims to discuss the generic parametrization of such spectrum in the presence of a nontrivial mass-renormalization in the Mukhanov-Sasaki action. As a consequence, the features and the interpretation of the divergent decoupling limit are not discussed.

 The paper is organized as follows. In section \ref{rg} we describe the theory and provide the RG formalism for the $\lq\lq\phi^2"$ potential. In section \ref{power spectra, free} we compute the tensor and scalar power spectra in the $m_{\chi}\to\infty$ limit, where the fakeon is decoupled from the theory. In section \ref{tensorQG} we compute the predictions in the full theory by projecting the fakeon from the physical spectrum and comment on the final results. Section \ref{Interpretation} is devoted to the physical interpretation of the predictions. Useful formulas are collected in the Appendix.

\section{Cosmic RG\ Flow for Quadratic Inflation}\label{rg}

We start by considering the classical action of the theory in the \textit{inflaton framework} (also known as \textit{Einstein frame})
 \begin{equation}\label{Sinfl}
     S_{\text{infl}}=-\frac{1}{16 \pi G}\int \sqrt{-g}\,\,d^4 x \left( R+\frac{1}{2 m_{\chi}^2} C^{\mu\nu\rho\sigma}C_{\mu\nu\rho\sigma}\right)
    + \frac{1}{2}\int \sqrt{-g}\,\,d^4 x \left( D_{\mu}\phi D^{\mu}\phi-2V(\phi)\right),
  \end{equation}
where $C_{\mu\nu\rho\sigma}$ is the Weyl tensor, while $m_{\phi}$ and $m_{\chi}$ are the inflaton and fakeon masses. In general, the choice of the potential $V(\phi)$ leads to a different inflationary dynamics. In particular, if we choose $V(\phi)$ to be the \textit{Starobinsky potential} \cite{starobinsky, vilenkin}, and perform the appropriate field transformations, we retrieve the $R+R^2+C^2$ action in the \textit{geometric framework} (or \textit{Jordan frame}) \cite{ABP}. 

As stated in the introduction, we focus on the model of quadratic inflation
\begin{equation}
    V(\phi)=\frac{m_{\phi}^2}{2}\phi^2.
\end{equation}
Despite the choice of a different potential from the Starobinsky one, we still have nontrivial phenomenological features due to the $C^2$ term, which is associated to the propagation of the purely virtual particles.

As a first step, we start with the background physics, which is described by the Friedmann-Lamaitre-Robertson-Walker (FLRW) metric
\begin{equation}
    g_{\mu\nu}=\text{diag}(1,-a^2,-a^2,-a^2).
\end{equation}
Here $a(t)$ is the expansion parameter.
Since the FLRW metric has a vanishing Weyl tensor, the Friedmann equations derived from (\ref{Sinfl}) are the ones of the usual slow roll inflation. In particular, we have

\begin{equation}\label{Friedeq}
    \dot{H}=-4\pi G\dot{\phi}^2,\,\,\,\,\,H^2=\frac{8\pi G}{3}\left(\frac{\dot{\phi}^2}{2}+V(\phi)\right),\,\,\,\,\,\,\ddot{\phi}+3H\dot{\phi}+V'(\phi)=0,
\end{equation}
where $H\equiv \dot{a}/a$ denotes the Hubble parameter.
We introduce the cosmic RG flow \cite{CMBrunning, HighOrder, Last} by defining the \textit{coupling constant}
\begin{equation}\label{coup}
    \alpha\equiv\sqrt{-\frac{\dot{H}}{3H^2}}=\sqrt{\frac{4\pi G}{3}}\frac{\dot{\phi}}{H}=\sqrt{\frac{\epsilon}{3}},
\end{equation}
which parametrizes the departure from the de Sitter universe in terms of the slow roll parameter $\epsilon \equiv -\dot{H}/H^2$. The insertion of (\ref{coup}) in the first two equations (\ref{Friedeq}) allows us to compute $\dot{\phi}$ and V($\phi$) in terms of the coupling. We obtain

\begin{equation}\label{field, pot}
    \dot{\phi}=\sqrt{\frac{3}{4\pi G}}\alpha H,\,\,\,\,\,\,\,\,\,\,\,\, V(\phi)=\frac{3}{8\pi G}(1-\alpha^2)H^2.
\end{equation}
The second equation sets the bound $-1\le\alpha\le 1$ since V($\phi$) is positive. Now we can insert (\ref{field, pot}) into the third equation (\ref{Friedeq}) to get the evolution equation of $\alpha$

\begin{equation}\label{evoalpha}
\dot \alpha = 3H\alpha(\alpha^2-1)-m_{\phi}\sqrt{1-\alpha^2}.
\end{equation}
Since we are dealing with an inflationary scenario we may adopt a perturbative approach: $\alpha$ is a small parameter. Then we can write the Hubble parameter as a power series in $\alpha$

\begin{equation}\label{Hexp}
   H=\sum_{n=0}^{\infty}h_n\alpha^n ,
\end{equation}

\noindent and  can determine the coefficients of the expansion by taking the time derivative of (\ref{Hexp}): in the LHS we use $\dot{H}=-3\alpha^2 H^2$, while in the RHS we replace $\dot{\alpha}$ with (\ref{evoalpha}). This procedure allows to extract  the $h_n$ coefficients order by order in the perturbative expansion. In particular, we have

\begin{equation}\label{Hexp, sing}
    \frac{m_{\phi}}{H}=-3\alpha-\frac{3}{2}\alpha^3+\frac{15}{8}\alpha^5-\frac{183}{16}\alpha^7+\mathcal{O}(\alpha^8).
\end{equation}
We note that the expression of $H$ is singular in the de Sitter limit ($\alpha=0$): in section \ref{Interpretation}, we  show that the appearance of this essential singularity produces physical effects when the $\phi^2$ model is combined with the $R+C^2$ action.
 Next, we switch to the conformal time
\begin{equation}
    \tau\equiv-\int_{t}^{+\infty}\frac{dt'}{a(t')}.
\end{equation}
Precisely, we consider the quantity $-a H\tau$ and write an analogous power expansion (\ref{Hexp}). Following the previous procedure, the outcome is 

\begin{equation}\label{aht}
    -a H \tau=1+3\alpha^2+27 \alpha^4+\frac{1573}{4}\alpha^6+\mathcal{O}(\alpha^8).
\end{equation}
On the other hand we can write $\frac{d}{dt}=-\frac{H}{(-aH\tau)}\frac{d}{d\ln{|\tau|}}$ and read from (\ref{evoalpha}) the \textit{$\beta$ function} of the theory,
\begin{equation}\label{betafull}
\beta_{\alpha}\equiv \frac{d \,\alpha(-\tau)}{d\ln{|\tau|}}=-(-a H\tau)\left[3\alpha(\alpha^2-1)-\frac{m_{\phi}}{H}\sqrt{1-\alpha^2}\right].
\end{equation}
Plugging the expansions (\ref{Hexp}) and (\ref{aht}), we obtain the perturbative expression 
 
\begin{equation}\label{beta}
\beta_{\alpha}=-3\alpha^3(1+2\alpha^2+28\alpha^4+\mathcal{O}(\alpha^6)).
\end{equation}
This is a QCD-like beta function. Specifically, we learn that $\alpha$=0 is the unique fixed point, and the theory is asymptotically free (de Sitter universe in the infinite past $\tau \rightarrow -\infty$). The uniqueness of the fixed point can be proved  by searching for zeros of (\ref{betafull}) of the form $\alpha$=const. This implies $\frac{d}{dt}\left(\frac{m_{\phi}}{H}\right)=0$ and therefore $\epsilon=3\alpha^2=0$. 

 We now move to the \textit{running} of the coupling constant $\alpha$. In particular, we switch to the dimensionless variable $\eta\equiv-k\tau$, where $k=|\mathbf{k}|$ is a reference comoving momentum and read the running equation from (\ref{beta})
    
    \begin{equation}\label{invbeta}
        d\ln{\eta}=-\frac{d\alpha}{3\alpha^3(1+2\alpha^2+28\alpha^4+\mathcal{O}(\alpha^6))}.
    \end{equation}
Finally, we can cast the running coupling in terms of a leading and subleading log expansion
\begin{equation}\label{alpha, nnll}
    \alpha^2=\frac{\alpha_k^2}{\lambda}\prod_{i = 0}^{2}(1+\alpha_k^{2n} \gamma_n(\lambda)),
\end{equation}
where $\alpha_k\equiv \alpha(k^{-1})$ and $\lambda=1+6\alpha_k^2 \ln{\eta}$, while $\gamma_n(\lambda)$ are functions that are determined by inserting (\ref{alpha, nnll}) into (\ref{invbeta}). With this procedure we can extract the functions $\gamma_n$ iteratively order by order. We give the results  for $i<3$ (LL, NLL, NNLL contributions)
\begin{equation}\label{running}
    \alpha^2=\frac{\alpha_k^2}{\lambda}\prod_{i = 0}^{2}(1+\alpha_k^{2n} \gamma_n(\lambda)),
\end{equation}
where 

\begin{equation}
    \begin{split}
\gamma_0(\lambda)&= 0 \\
\gamma_1(\lambda) &= -\frac{4 \ln{\lambda}}{\lambda}\\
\gamma_2(\lambda) &= -\frac{12}{\lambda^2}\left(1-\lambda+\frac{4}{3}\ln{\lambda} (1-\ln{\lambda})\right).
\end{split}
\end{equation}

As discussed in \cite{Last}, the employed formalism exhibits a RG structure in the proper sense. Indeed, the resulting correlation functions (i.e. the power spectra, see section \ref{power spectra, free}) satisfy an equation of the Callan-Symanzik type  with vanishing anomalous dimension \cite{CMBrunning}. 

Having discussed the main features of the cosmic RG flow, we can study the fluctuations around the background metric and their associated power spectra.

\section[Limit of heavy fakeon]{Power Spectra in the
infinitely heavy fakeon limit}\label{power spectra, free}
\setcounter{equation}{0}
In this section we compute the power spectra in the heavy fakeon limit by means of the cosmic RG formalism. In particular, we recover the known results in the literature \cite{Last} and further extend the de Sitter perturbative expansion.

The action of the inflaton framework in the $m_{\chi}\to\infty$ limit reads
\begin{equation}\label{fakeon free}
    \begin{split}
    S_{\text{infl}}&=-\frac{1}{16 \pi G}\int \sqrt{-g}\,\,d^4 x  R
    + \frac{1}{2}\int \sqrt{-g}\,\,d^4 x \left( D_{\mu}\phi D^{\mu}\phi-2V(\phi)\right),
  \end{split}
\end{equation}
so that  we retrieve  General Relativity coupled to the inflaton sector.
\subsection{Tensor modes}
 Let us parametrize the perturbations associated to the tensor modes in the following way \cite{reviews}

\begin{equation}\label{gTensor}
\begin{split}
    g_{\mu\nu}=(1,-a^2,-a^2,-a^2)-2a^2(u\delta_{\mu}^1\delta_{\nu}^1-u\delta_{\mu}^2\delta_{\nu}^2+v\delta_{\mu}^1\delta_{\nu}^2+v\delta_{\mu}^2\delta_{\nu}^1),
\end{split}
\end{equation}

\noindent where $u(t,z)$ and $v(t,z)$ are the physical graviton polarizations\footnote{$u$ and $v$ depend only on the $z$ spatial coordinates because the graviton polarizations are helicity eigenstates.}. We now insert (\ref{gTensor}) into the action (\ref{fakeon free}) and express the modes via their spatial Fourier transform ($u_{\mathbf{k}}(t)$ and $v_{\mathbf{k}}(t)$). We get, up to the quadratic order in perturbation theory, the following action

\begin{equation}
    S_t=\int dt d^3k\frac{a^3}{8\pi G}\left[\dot u_{\mathbf{k}}(t) \dot u_{-\mathbf{k}}(t)-\frac{k^2}{a^2} u_{\mathbf{k}}(t)u_{-\mathbf{k}}(t)\right]+\text{same for $v_{\mathbf{k}}$}.
\end{equation}
In particular, we derived this result working in the \textit{comoving gauge} $\delta\phi=0$ and plugging (\ref{field, pot}) in (\ref{fakeon free}).

Let us focus on the $u_{\mathbf{k}}$ mode. We perform the change of variable\footnote{In what follows, we often omit the $\mathbf{k}$ subscript for compactness.}
\begin{equation}\label{u to w}
    w=a u\sqrt{\frac{k}{4\pi G}}.
\end{equation}
Thanks to this redefinition, the $a^3$ dependence in front of the kinetic term vanishes when we switch to the conformal time $\eta$. What we get, upon integrating by parts and the insertion of (\ref{aht}), is

\begin{equation}\label{w action}
    S_t=\frac{1}{2}\int d\eta d^3k\left[w'^2-w^2+(2+\sigma_t)\frac{w^2}{\eta^2}\right]
\end{equation}
where the prime denotes the derivative with respect to $\eta$, while 
\begin{equation}\label{sigt}
    \sigma_t= 9 \alpha^2+108 \alpha^4+1708 \alpha^6+\mathcal{O}(\alpha^8). 
\end{equation}
    The corresponding equation of motion (\textit{Mukhanov-Sasaki} equation) is
\begin{equation}\label{eom wt}
    w''+w-(2+\sigma_t)\frac{w}{\eta^2}=0.
\end{equation}

Following the standard procedure, we quantize the metric perturbations by promoting the $w_{\mathbf{k}}$ modes to operators: $\hat{w}_{\mathbf{k}}(\eta)=w_{\mathbf{k}}(\eta)\hat{a}_{\mathbf{k}}+w_{-\mathbf{k}}^*(\eta)\hat{a}_{-\mathbf{k}}^{\dagger}$, where $a_{\mathbf{k}}$ and $a_{-\mathbf{k}}^{\dagger}$ are the usual creation and annihilation operators. Then, the vacuum state of this quantum theory is fixed by setting the \textit{Bunch-Davies  condition} \cite{BD vacuum}
\begin{equation}\label{BD}
 w\sim \frac{e^{i\eta}}{\sqrt{2}},\,\,\,\,\,\,\,\,\eta\to\infty.
\end{equation}

We now solve (\ref{eom wt}) equipped with (\ref{BD}). In particular, we can write the $w$ modes as $w=\sum_{n=0}^{\infty}w_n\alpha_k^n$, so that (\ref{eom wt}) reads

\begin{equation}\label{w, g_n}
    w_n''+w_n-\frac{2w_n}{\eta^2}=\frac{g_n^t(\eta)}{\eta^2}
\end{equation}
order by order in perturbation theory. The $g_n^t(\eta)$ are known functions and are listed in the Appendix. Ref. \cite{CMBrunning} shows that is possible to write the $w$ mode as two contributions: $\eta w=Q(\ln{\eta})+W(\eta)$. The first contribution is dominant in the super-horizon limit ($\eta \to 0$), while the second one is vanishing. Specifically, the following \textit{Q-equation} holds \cite{CMBrunning, HighOrder}
\begin{equation}\label{Qeq}
   \beta_{\alpha}\frac{\partial \widetilde{Q}}{\partial \alpha}=-\frac{\sigma}{3}\widetilde{Q}-\frac{1}{3}\sum_{n=1}^{\infty}3^{-n}\left(\beta_{\alpha}\frac{\partial}{\partial\alpha}\right)^n(\sigma \widetilde{Q}),
    \end{equation}
where $\widetilde{Q}(\alpha,\alpha_k)\equiv Q(\ln{\eta}(\alpha,\alpha_k))$.
We now solve equation (\ref{Qeq}) up to the NNLL order. In particular, we truncate $\beta_{\alpha}$ and $\sigma_t$ as in equations (\ref{beta}), (\ref{sigt}), while the RHS of (\ref{Qeq})  contains the first three terms
\begin{equation}\label{Qeq, NNLL}
    \beta_{\alpha}\frac{\partial \widetilde{Q}}{\partial \alpha}=-\frac{\sigma}{3}\widetilde{Q}-\frac{1}{9}\left(\beta_{\alpha}\frac{\partial}{\partial\alpha}\right)(\sigma \widetilde{Q})-\frac{1}{27}\left(\beta_{\alpha}\frac{\partial}{\partial\alpha}\right)\left[\left(\beta_{\alpha}\frac{\partial}{\partial\alpha}\right)(\sigma \widetilde{Q})\right].
\end{equation}
We can seek a solution for $\widetilde{Q}$ in power series

\begin{equation}\label{Q sol}
     \widetilde{Q}(\alpha,\alpha_k)=\widetilde{Q}(\alpha_k)\frac{\alpha}{\alpha_k}\frac{1+\sum_{n=1}^{\infty}c_n\alpha^{2n}}{1+\sum_{n=1}^{\infty}c_n\alpha_k^{2n}}.
 \end{equation}
$\widetilde{Q}(\alpha_k)$ is an arbitrary constant that must be fixed by means of the Bunch-Davies condition. We determine the $c_n$ coefficients by inserting (\ref{Q sol}) into (\ref{Qeq, NNLL}), which gives 
\begin{equation}\label{ansatz, tensor}
      c_1=\frac{7}{2},\,\,c_2=\frac{2057}{72}.
  \end{equation}
In deriving the $c_n$ coefficients, we have also used the running coupling (\ref{running}).

After the determination of the $w$ modes in the super-horizon limit, we can compute the tensor power spectrum. The tensor power spectrum $\mathcal{P}_{T}$ is defined from the two-point correlator\footnote{We have already summed the (equal) contributions of the $u_{\mathbf{k}}$ and $v_{\mathbf{k}}$ polarizations in  formula (\ref{def pt}).}

\begin{eqnarray}\label{def pt}
\langle \hat{u}_{\mathbf{k}}(\tau )\hat{u}_{\mathbf{k}^{\prime }}(\tau
)\rangle  &=&(2\pi )^{3}\delta ^{(3)}(\mathbf{k}+\mathbf{k}^{\prime })\frac{%
\pi ^{2}}{8k^{3}}\mathcal{P}_{T},\qquad \mathcal{P}_{T}=\frac{8k^{3}}{\pi
^{2}}|u_{\mathbf{k}}|^{2},  \label{pt} 
\end{eqnarray}
and it has the remarkable property of being time-independent on super-horizon scales \cite{Baumann, reviews}. Starting from the solution (\ref{Q sol}), we now rewind all the field redefinition and plug them in (\ref{pt}). In particular, we extract the $w$ modes via $w\sim \widetilde{Q}/\eta$ and use (\ref{u to w}) combined with the power expansion (\ref{aht}). The result is
 
 \begin{equation}
     \mathcal{P}_T=\frac{32Gm_{\phi}^2}{9\pi\alpha_k^2}|\widetilde{Q}(\alpha_k)|^2\left[1-7\alpha_k^2+\mathcal{O}(\alpha_k^4)\right],
 \end{equation}
 which is correctly time independent.

 The next step is to fix the $\widetilde{Q}(\alpha_k)$ constant by means of the Bunch-Davies condition. Precisely, we solve (\ref{w, g_n}) with  the functions $g_n^t$ collected in the Appendix and fix the arbitrary constants of each solution via Bunch-Davies\footnote{Of course the Bunch-Davies condition is imposed on the whole $w$ mode. This requirement is \lq\lq transferred" on each $w_n$ by imposing $w_0\sim\frac{e^{i\eta}}{\sqrt{2}},w_1\sim 0, w_2\sim0$, $w_3\sim 0$... as $\eta\rightarrow\infty$.}. Finally, we sum these contributions by imposing

\begin{equation}
    \widetilde{Q}(\alpha_k)=\lim_{\eta\rightarrow 0}\eta w \Big\rvert_{\alpha=\alpha_k} =\lim_{\eta\rightarrow 0}\eta[w_0+w_1\alpha+w_2\alpha^2+w_3\alpha^3]\Big\rvert_{\alpha=\alpha_k}.
\end{equation}
The four $w_n$ functions are listed in the Appendix. The outcome is
 
 \begin{equation}
    \widetilde{Q}(\alpha_k)=\frac{i}{\sqrt{2}}\left[1+3(2-\gamma_M+i\pi)\alpha_k^2+\mathcal{O}(\alpha_k^4)\right],
 \end{equation}
where $\gamma_M$ is the Euler-Mascheroni constant and the final tensor power spectrum reads

 \begin{equation}
     \mathcal{P}_T=\frac{16G m_{\phi}^2}{9\pi\alpha_k^2}\left[1+(5-6\gamma_M)\alpha_k^2+\mathcal{O}(\alpha_k^4)\right].
 \end{equation}
This final result agrees with the general results of Ref. \cite{Last}.

\subsection{Scalar Modes}
We parametrize the scalar perturbations of the metric as follows \cite{reviews}

\begin{equation}\label{metric, scalar}
    g_{\mu\nu}=(1,-a^2,-a^2,-a^2)+2(\Phi,a^2\Psi,a^2\Psi,a^2\Psi)-\delta_{\mu}^0\delta_{\nu}^i\partial_iB-\delta_{\nu}^0\delta_{\mu}^i\partial_iB.
\end{equation}
We insert (\ref{metric, scalar}) in the action (\ref{fakeon free}) by working in the comoving gauge $\delta\phi=0$. In terms of the Fourier components, the corresponding lagrangian is

\begin{equation}\label{scalar L}
    \frac{8\pi G}{a^3}\mathcal{L}_s=-3(\dot{\Psi}+H\Phi)^2+4\pi\dot{\phi}^2\Phi^2+\frac{k^2}{a^2}\left[2B(\dot{\Psi}+H\Phi)+\Psi(\Psi-2\Phi)\right].
\end{equation}
Using the background value for $\dot{\phi}$ and noting that $B$ appears algebraically in (\ref{scalar L}), we  integrate out $B$ (its equation of motion implies $\Phi=-\frac{\dot{\Psi}}{H}$) and obtain

\begin{equation}
    \frac{8\pi G}{a^3}\mathcal{L}_s=3\alpha^2\left(\dot{\Psi}^2-\frac{k^2}{a^2}\Psi^2\right).
\end{equation}
 We can easily obtain the Mukhanov-Sasaki action by performing the following field redefinition

\begin{equation}
    w\equiv\alpha a\Psi \sqrt\frac{3k}{4\pi G}.
\end{equation}
 In this way, we obtain the same action (\ref{w action})  in the conformal time 

\begin{equation}
    \mathcal{L}_s=\frac{1}{2}\left[w'^2-w^2+(2+\sigma_s)\frac{w^2}{\eta^2}\right],
\end{equation}
with $\sigma_s=18\alpha^2+171\alpha^4+2302\alpha^6+\mathcal{O}(\alpha^8)$. 

The computation of the $\Psi_{\mathbf{k}}$ mode is identical to the one of $u_{\mathbf{k}}$. Precisely, we first quantize the fluctuations and then solve the  NNLL \textit{Q equation} (\ref{Qeq, NNLL}) for $\sigma_s$  with the following ansatz in power series

\begin{equation}
   \widetilde{Q}(\alpha,\alpha_k)=\widetilde{Q}(\alpha_k)\frac{\alpha^2}{\alpha^2_k}\frac{1+\sum_{n=1}^{\infty}c_n\alpha^{2n}}{1+\sum_{n=1}^{\infty}c_n\alpha_k^{2n}}.  
 \end{equation}
We find that the $c_n$ coefficients are the same as those of equation (\ref{ansatz, tensor}).

We now can compute the scalar power spectrum $\mathcal{P}_{\mathcal{R}}$, which is defined from the two-point correlation function of the \textit{comoving curvature perturbation} $\mathcal{R}$ (which is equal to $\Psi$ in our gauge choice \cite{Baumann}),
\begin{eqnarray}
\langle \mathcal{R}_{\mathbf{k}}(\tau )\mathcal{R}_{\mathbf{k}^{\prime
}}(\tau )\rangle  &=&(2\pi )^{3}\delta ^{(3)}(\mathbf{k}+\mathbf{k}^{\prime
})\frac{2\pi ^{2}}{k^{3}}\mathcal{P}_{\mathcal{R}},\qquad \mathcal{P}_{%
\mathcal{R}}=\frac{k^{3}}{2\pi ^{2}}|\Psi _{\mathbf{k}}|^{2}.  \label{pR}
\end{eqnarray}

 Finally, using the derived power expansions, (\ref{pR}) gives

\begin{equation}
    \mathcal{P}_{\mathcal{R}}=\frac{2Gm_{\phi}^2}{27\pi\alpha_k^4}|\widetilde{Q}(\alpha_k)|^2\left[1-7\alpha_k^2+\mathcal{O}(\alpha_k^4)\right].
\end{equation}
 The $g_n^s$ and $w_n^s$ functions for the scalar case are reported in the Appendix. Here we just give the result of $\widetilde{Q}(\alpha_k)$ obtained by imposing the Bunch-Davies condition, which reads

\begin{equation}
  \widetilde{Q}(\alpha_k)=\frac{i}{\sqrt{2}}\left[1+(12-6\gamma_M+6i\pi)\alpha_k^2+\mathcal{O}(\alpha_k^4)\right].
 \end{equation}
The final scalar power spectrum is
\begin{equation}\label{Pr}
     \mathcal{P}_{\mathcal{R}}=\frac{Gm_{\phi}^2}{27\pi\alpha_k^4}\left[1+(17-12\gamma_M)\alpha_k^2+\mathcal{O}(\alpha_k^4)\right].
 \end{equation} 
As before, the power spectrum is time independent in the super-horizon limit. The leading term in the expansion of (\ref{Pr}) coincides with the general result of Ref. \cite{Last}: in this paper, we also provide the first nontrivial subleading correction.

Finally, we introduce the tensor to scalar ratio and the tensor and scalar spectral indices in terms of the $\beta$ function \cite{CMBrunning}
\begin{equation}\label{r, n}
     \begin{split}
         r(k)&=\frac{\mathcal{P}_T}{\mathcal{P}_{\mathcal{R}}}\\
         n_{T}&=-\beta_{\alpha}(\alpha_k)\frac{\partial \ln{\mathcal{P}_T}}{\partial \alpha_k},\,\,\, n_{\mathcal{R}}-1=-\beta_{\alpha}(\alpha_k)\frac{\partial\ln{ \mathcal{P}_{\mathcal{R}}}}{\partial \alpha_k}.
     \end{split}
 \end{equation}
In our specific case, we get
\begin{equation}\label{pred}
    r=48\alpha_k^2+\mathcal{O}(\alpha_k^4),\,\,\,\,\, n_{T}=-6\alpha_k^2+\mathcal{O}(\alpha_k^4),\,\,\,\,\,n_{\mathcal{R}}-1=-12\alpha_k^2+\mathcal{O}(\alpha_k^4),
\end{equation}
 which yields  to the well known consistency relation
\begin{equation}
     r+8n_T=\mathcal{O}(\alpha_k^4).
 \end{equation}
As a further check of the formalism, we combine (\ref{pred}) with the Planck data \cite{Planck18}. In particular, we obtain $\alpha_{0.002}=0.053\pm0.005$ from the measured $n_{\mathcal{R}}$ at the pivot scale $k=0.002\,\text{Mpc}^{-1}$. This value in turn yields to $r_{0.002}=0.136\pm 0.016$, which is largely excluded by the current upper bound $r_{0.002}< 0.055$. Consistently to what is known in the literature, we find that the $\lq\lq\phi^2"$ slow roll potential combined with General Relativity is excluded by Planck data.
 
\section{Tensor Modes from Purely Virtual Particles and the Singular Decoupling Limit}

\label{tensorQG}

In the following section we discuss the inclusion of purely virtual particles in quadratic inflation. As a prototype computation, we derive the tensor power spectrum by projecting away the purely virtual particle from the physical spectrum. In the following analysis, we exclusively deal with the tensor modes: all the nontrivial aspects of the $\lq\lq\phi^2"$ case are already encoded here. 

Parametrizing the metric as in (\ref{gTensor}), we find the following lagrangian $\mathcal{L}_t$
\begin{equation}
\begin{split}
    \frac{8 \pi G}{a^3} \mathcal{L}_t&=\left(\dot u^2(t) -\frac{k^2}{a^2} u^2(t)\right)+
   \\
    &-\frac{1}{ m_{\chi}^2}\left[\ddot  u^2(t)-2\left(H^2-\frac{3}{2}\alpha^2 H^2+\frac{k^2}{a^2} \right)\dot u^2(t)+\frac{k^4}{a^4} u^2(t)\right].
    \end{split}
\end{equation}
The presence of the $C^2$ term in (\ref{Sinfl}) introduces a high-derivative term in the fluctuations. We can remove this term by adding an auxiliary lagrangian containing a new algebraic field $U(t)$
\begin{equation}
    \Delta \mathcal{L}_t=\frac{a^3}{8 \pi G m_{\chi}^2} \left(m_{\chi}^2 c(t) U - \ddot u -
    f(t) \dot u- h(t) u\right)^2,
\end{equation}
where $f,c,h$ are arbitrary functions that are  chosen in order to cast the lagrangian in the simplest form possible. Precisely, by means of the field redefinition 
\begin{equation}
    u=U+V,
\end{equation}
we can cast the lagrangian in the form
\begin{equation}\label{lagr , t}
\begin{aligned}
 \frac{8 \pi G}{a^3 \gamma} \mathcal{L}_t=\dot{U}^2-\left[\frac{k^2 }{a^2}+g_U(\gamma,H,\alpha,k)\right]U^2 - \dot{V}^2 +\\
+\left[\frac{k^2}{a^2}+g_V(\gamma,H,\alpha,k)\right]V^2+g_{UV}(\gamma,H,\alpha,k)UV,
 \end{aligned}
\end{equation}
 with the following choice of the arbitrary functions
\begin{align}  \label{funct}
\begin{split}
c(t) &= \gamma \equiv 1+2\frac{H^2}{m_{\chi}^2},\\
f(t) &= 3H-\frac{12 \alpha^2 H^3 }{m_{\chi}^2 \gamma},\\
h(t) &=\frac{k^2}{a^2}+m_{\chi}^2\gamma+\\&-3H^2\alpha\left[\frac{4 m_{\phi}\,H}{m_{\chi}^2}\sqrt{1-\alpha^2}-\left(1-\frac{18H^2}{m_{\chi}^2}\right)\alpha+\left(1-\frac{8H^2}{m_{\chi}^2\gamma}\right)\frac{6\alpha^3H^2}{m_{\chi}^2\gamma}\right].
\end{split}
\end{align}
While $g_{U,V,UV}$ are known (but involved) functions that we do not report here.\\
\indent Let us consider the field redefinitions
\begin{equation}
\begin{aligned}
U &\equiv \sqrt{\frac{4\pi G}{\gamma}}\frac{U_1}{a},\\
V &\equiv  \sqrt{\frac{4 \pi G}{\gamma}}\frac{V_1}{a}.
\end{aligned}
\end{equation}
Now it is possible to cancel the $a$ and $\gamma$ dependence in front of the kinetic terms by  switching  to the conformal time $\tau$. The lagrangian reads

\begin{equation}\label{unproj}
\begin{aligned}
\mathcal{L}_t=\frac{1}{2}U_1'^2-\frac{1}{2}\left(k^2 +\mathcal{F}_U(\gamma,H,\alpha,k,a)\right)U_1^2 - \frac{1}{2} V_1'^2 +\\
+\frac{1}{2}\left(k^2+\mathcal{F}_V(\gamma,H,\alpha,k,a)\right)V_1^2+\mathcal{F}_{UV}(\gamma,H,\alpha,k,a)U_1V_1,
 \end{aligned}
\end{equation}
 where $\mathcal{F}_{U,UV,V}$, again, are known functions. Every expression until now is \textit{exact} in $\alpha$. Furthermore, the heavy fakeon limit $ m_{\chi}\rightarrow \infty$ of (\ref{unproj}) retrieves correctly the $\sigma_t$ expression derived in the fakeon-free theory (\ref{sigt}). The next step is to expand (\ref{unproj}) around the de Sitter fixed point by using (\ref{Hexp, sing}) and (\ref{aht}). Remarkably, up to $\alpha^4$ order, we get

\begin{equation}\label{finalL}
\begin{aligned}
\mathcal{L}_t=\frac{1}{2}U_1'^2-\frac{1}{2}\left(b(\alpha)\, k^2 -\frac{1}{\tau^2}\left[2+\left(18+\frac{81}{2\xi^2}\right)\alpha^4\right]\right)U_1^2 - \frac{1}{2} V_1'^2 +\\
+\frac{1}{2}\left((2-b(\alpha))k^2+\frac{\alpha^2}{\tau^2}\left[\frac{9}{\xi^2}+\left(18+\frac{207}{2\xi^2}\right)\alpha^2\right]\right)V_1^2+\\
+ \left((1-b(\alpha))k^2-\frac{\alpha^2}{\tau^2}\left[6+\left(45-\frac{81}{2\xi^2}\right)\alpha^2\right]\right)U_1V_1,
 \end{aligned}
\end{equation}
 where $\xi \equiv \frac{m_{\phi}}{m_{\chi}}$, and 
\begin{equation}
    b(\alpha)=1-\left(36+\frac{27}{\xi^2}\right)\alpha^4
\end{equation}
is a sort of \textit{running squared mass}, similar to what we get in the case of the Starobinsky potential with purely virtual particles. We note three features from (\ref{finalL}):
\begin{itemize}
    \item The particular choice of the arbitrary functions (\ref{funct}) has the advantage of admitting a smooth de Sitter limit.
    \item The lagrangian is diagonal in the de Sitter limit and $V_1(\tau)$ is indeed the fakeon, as can be seen from the negative kinetic term.
    \item The heavy fakeon limit $\xi\rightarrow0$ no longer retrieves the fakeon free-theory.
\end{itemize}
We emphasize that the third feature could be regarded (at least at this stage) just as a drawback of the chosen parametrization\footnote{Other choices were studied to overcome this unpleasant feature but, unfortunately, they lead to singular expressions in the de Sitter limit (which is a much more annoying issue for computations). }: in order to understand if this is a real issue, we have to carry out  the computation of the tensor power spectrum.

\subsection{The Fakeon Projection}\label{fakeon projection}

 At the classical level, the fakeon projection amounts to replace the solution of the $V_1$ equation of motion (calculated with the fakeon prescription \cite{classicization, FLRW}) into the lagrangian (\ref{finalL}). The $U_1V_1$ term is of order $\alpha^2$, therefore, we expect the fakeon to be of order $\alpha^2$ as well. In this case (at the cubic order in $\alpha$), the Bunch-Davies condition is not affected by $V_1$. Therefore, we can consider the super-horizon limit $k\tau\rightarrow 0$ for the fakeon equation of motion. We determine the solution  by means of the following \textit{ansatz} in power expansion:
 \begin{equation}\label{ansatz}
     V_1=\alpha^2(b_0+b_1\alpha^2)U_1+b_{2}\alpha^4 \tau U_1'.
 \end{equation}
 The coefficients are determined as follows: when we insert the ansatz in the fakeon equation of motion there are terms such as $U_1''$ and $U_1^{(3)}$. These terms can be replaced by means of the $U_1$ equation of motion using again the \textit{ansatz} (\ref{ansatz}) when necessary. At this point, exploiting the super-horizon limit $k \tau\rightarrow0$,  we obtain an equation of the form
\begin{equation}
    f_1({b_i},\alpha) U_1+f_2({b_i},\alpha)\tau U_1'=0,
\end{equation}
where $f_{1,2}$ are power expansions in $\alpha$. Setting each term of the power expansion to zero, we get the explicit values of the coefficients, which are
\begin{equation}\label{coeff}
    b_0=3,\,\,\,b_1=\frac{27}{2}\left(1-\frac{5}{2\xi^2}\right),\,\,\,b_2=18.
\end{equation}
\indent It is important to stress that the ansatz in power series captures the perturbative behavior of the full fakeon Green function $G_f(t,t')$ (obtained through the classical limit of the quantum prescription, see the next section). We thus see  that the insertion of the fakeon solution into the lagrangian (\ref{finalL}) leads to non-local terms, due to the presence of the full fakeon Green function. These terms, however, are of order $\alpha^4$, since $V_1$ is of order $\alpha^2$: in other words, the action is still local up to the $\alpha^3$ order\footnote{In this case we cannot overcome the issue of non locality by using the $\xi\rightarrow 0$ expansion as in the scalar case of the Starobinsky potential \cite{CMBrunning} because every expression is singular in this limit.}.

\subsection{Tensor power spectrum}

Inserting the \textit{ansatz} in (\ref{ansatz}), we obtain (up to $\alpha^3$)

\begin{equation}
\mathcal{L}_{proj}=\frac{1}{2}U_1'^2-\frac{1}{2}\left(k^2-\frac{2}{\tau^2}\right)U_1^2 ,
\end{equation}
 which coincides with the lagrangian in the de Sitter case. We can scale the field $U_1\rightarrow U_2\equiv \sqrt{k} U_1$ (coherently, we scale $V_1$ as well) and switch to the dimensionless variable $\eta=-k\tau$. After the quantization of the field, the solution of the equation of motion equipped with the Bunch-Davies condition reads
\begin{equation}\label{mode desitter}
    U_2=\frac{e^{i\eta}}{\sqrt{2}}\left(1+\frac{i}{\eta}\right).
\end{equation}
Now we can rewind the changes of variables and compute the tensor power spectrum $\mathcal{P}_T$. The result is

\begin{align}\label{pt, fake}
\begin{split}
 \mathcal{P}_T &=\frac{8k^3}{\pi^2}|u_{\mathbf{k}}|^2=\frac{8k^3}{\pi^2}|U+V|^2=\\
     &= \frac{32 k^3 G}{\pi\gamma a^2}|U_1+V_1|^2=\\
     &= \frac{32  G}{\pi}\frac{1}{(a H\tau)^2}\left(\frac{H^2}{1+\frac{2H^2}{m_{\chi}^2}}\right)\eta^2|U_2+V_2|^2.
     \end{split}
\end{align}
Expanding around the de Sitter fixed point by means of (\ref{Hexp, sing}) ,(\ref{aht}), (\ref{ansatz}) and taking the \textit{super-horizon} limit ($\eta\rightarrow 0$) after the insertion of (\ref{mode desitter}) into (\ref{pt, fake}), we obtain

\begin{equation}\label{pt, fake1}
    \mathcal{P}_T=\frac{8G m_{\phi}^2}{\pi\xi^2}\left(1-\frac{9}{2\xi^2}\alpha_k^2+\mathcal{O}(\alpha_k^4)\right).
\end{equation}
This last expression shows that the \textit{physical observable}, i.e. the tensor power spectrum, does not admit a finite decoupling limit $\xi\to 0$.

It is interesting to compare the final results with those presented in the literature with alternative approaches. In Ref. \cite{salvio}, the power spectra are computed by keeping the spin-2 ghost. In particular, it is argued that the effects of the ghost particle vanish for super-horizon scales, since the ghost $w_{\mathbf{k}}$ modes are suppressed in this limit. However, we recall that the $w_{\mathbf{k}}$ modes are determined by imposing the Bunch-Davies condition, which is set in the UV regime ($\eta\rightarrow\infty$). There, we cannot neglect the effects of the ghost. On the other hand, the fakeon prescription is crucial to deal with the UV regime. For example, in the $R+R^2+C^2$ theory (Starobinsky potential) with fakeons \cite{ABP}, the UV-IR interpolation gives a bound that involves the fakeon mass and, consequently, the tensor to scalar ratio $r$. Interestingly, Ref.\cite{NewSalvio} shows that the results of the ghost approach coincide with the predictions of the theory with fakeons once the aforementioned consistency bound is applied. 
 
  In the case of the $\lq\lq\phi^2"$ potential, the expression (\ref{pt, fake}) coincides with the one of Ref.~\cite{salvio} for the leading order contribution. This happens because the fakeon Green function does not contribute to the computations up to the $\alpha^3$ order. However, as we show in the next section, the fakeon prescription hides an additional constraint, which is able to explain the unusual properties of (\ref{pt, fake1}). 
\section{Physical Interpretation of the Singular Decoupling Limit}\label{Interpretation}
\setcounter{equation}{0}

In this section we discuss the key features of the singular decoupling limit. We first study the problem by enlightening the connection with the causal structure of the theory, then we examine the viability of the model from a phenomenological point of view. In what follows, we make a comparison with the Starobinsky potential, which admits a smooth and finite fakeon decoupling.

 The problem discussed in the previous section originates from a sort of \lq\lq discontinuity" in the computation: every step of the exact computation reproduces the fakeon free results under the decoupling limit and only after the step concerning the de Sitter limit, we encounter the first singularities in $\xi=0$. From an algebraic point of view is easy to understand the origin of the singular decoupling limit. Consider for example the variable $\gamma^{-1}$ that multiplies the power spectrum $\mathcal{P}_T$ 
 
\begin{equation}
    \gamma^{-1}=\frac{1}{1+\frac{2H^2}{m_{\chi}^2}}.
\end{equation}

 The leading term of the power expansion in the de Sitter limit gives
 
 \begin{equation}
     \gamma^{-1}\simeq
     \begin{cases}\frac{2 m_{\chi}^2}{m_{\phi}^2+2m_{\chi}^2} & \mbox{for } \mbox{Starobinsky}, \\ 
     \frac{9}{2}\left(\frac{m_{\chi}}{m_{\phi}}\right)^2\alpha^2 & \mbox{for } \mbox{$\phi^2$ }.
    \end{cases}
 \end{equation}
 We have used the $H$ power expansions of the Starobinsky \cite{ABP} and $\lq\lq\phi^2"$ potentials. The upper expression admits a regular decoupling limit, while the lower expression does not: clearly the difference between the two cases is encoded in the singular behavior of $H$  for the $\lq\lq\phi^2"$ potential. 
 
 The irregularity of the decoupling limit is enforced by the coefficients of the fakeon solution (\ref{coeff}), because they were already singular in $\xi=0$, well before the computation of $\mathcal{P}_T$ in (\ref{pt, fake}). This last remark may prompt the following question: is the projection procedure hiding some physical information? Indeed, the procedure showed in section \ref{fakeon projection} is more involved then the simple \lq\lq hunt for the right ansatz" of the fakeon equation of motion. The fakeon solution should be derived using the average of the \textit{advanced} and \textit{retarded} Green functions \cite{classicization} and the power expansion (\ref{ansatz}) is only able to capture the perturbative behavior of the Green function. The complete procedure has been explicitly applied for the Starobinsky case to the lowest order in the de Sitter expansion \cite{ABP}.

 The full fakeon Green function is constrained to have a real pole mass (\textit{no-tachyon condition}): this feature is crucial to make the theory causal\footnote{At least for time scales larger than $\Delta t\sim 1/m_{\chi}\sim 10^{-38}\, s.$ The estimate is obtained by plugging the benchmark value $m_{\chi}\simeq 10^{13}$ GeV, which is suggested by the predicted band for the tensor to scalar ratio \cite{ABP}.}, since a tachyonic mass would propagate the causality violation introduced by the purely virtual particles \cite{Absograv, classicization, FLRW} up to large time scales. In the Starobinsky case, this condition imposes a constraint on the fakeon and inflaton masses such that it is possible to make the prediction on the tensor to scalar ratio sharp \cite{ABP, CMBrunning}. 
 
 We now extend the analysis to the $\phi^2$ potential, where we  meet some additional difficulties due to the singularities in $\alpha=0$. In order to get some physical information, we work with the cosmological time . In particular, we switch to the new variables
 
 \begin{equation}
     A\equiv\sqrt{\frac{a^3\gamma}{4\pi G}}U,\,\, B\equiv\sqrt{\frac{a^3\gamma}{4\pi G}}V.
 \end{equation}
In this way, the lagrangian (\ref{lagr , t}) reads
 
 \begin{equation}
     \mathcal{L}_t=\frac{1}{2}\dot{A}^2-\frac{1}{2}\dot{B}^2+\mathcal{G}_A(\alpha) A^2+\mathcal{G}_B(\alpha) B^2+\mathcal{G}_{AB}(\alpha)AB,
 \end{equation}
 where, as usual, $\mathcal{G}$ are known functions. The de Sitter limit of this expression has some unexpected features: first of all, the singularities in $\alpha=0$ do not vanish as happens in the conformal time expression of $\mathcal{L}_t$. Secondly, the lagrangian is no longer diagonal in the de Sitter limit. In particular, we have
 
 \begin{equation}\label{off diag}
     \mathcal{L}_t=\frac{1}{2}\dot{\mathcal{U}}^{T}\begin{bmatrix}
1 & 0 \\
0 & -1 
\end{bmatrix}\dot{\mathcal{U}}-\frac{1}{2}\mathcal{U}^{T} M^{(2)}\mathcal{U},
 \end{equation}
 
 where
 
 \begin{equation}\label{mass eigen}
     \mathcal{U}\equiv \begin{bmatrix}
A  \\
B  
\end{bmatrix}\,\,\, \mbox{and}\,\,\, M^{(2)}\equiv \begin{bmatrix}
\frac{1}{4}\left(7-\frac{1}{\alpha^2}\right)m^2_{\phi}  & \frac{2}{3}m^2_{\phi} \\
\frac{2}{3}m^2_{\phi}  & -m_{\chi}^2-\frac{1}{36} \left(7-\frac{1}{\alpha^2}\right)m_{\phi}^2.
\end{bmatrix}
 \end{equation}
 Another issue concerning the computation of the Green function is the following: the equation of motion has some  time dependent coefficients due to the presence of $\alpha$. This is not a real problem. In fact, we recall that the smallness of the second slow roll parameter $\delta\equiv -\ddot{\phi}/H\dot\phi$ ensures a very slow variation of the $\epsilon$ parameter (and therefore $\alpha$) during inflation. Therefore, we assume $\alpha\simeq$const for this discussion.\\
 \indent Let us diagonalize the lagrangian (\ref{off diag}) by switching to the mass eigenstates. In particular, we introduce a new field $\mathcal{U} \equiv S\mathcal{B}$, where $S=\begin{bmatrix}
\cosh{\theta} & \sinh{\theta} \\
\sinh{\theta} & \cosh{\theta} 
\end{bmatrix}$ is a hyperbolic rotation. The lagrangian (\ref{off diag}) in terms of the new field reads 
\begin{equation}
    \mathcal{L}_t=\frac{1}{2}\dot{\mathcal{B}}^T\begin{bmatrix}
1 & 0 \\
0 & -1 
\end{bmatrix}\dot{\mathcal{B}}-\frac{1}{2}\mathcal{B}^TM_d^{(2)}\mathcal{B},
\end{equation}
where 

\begin{equation}\label{mass}
    M_d^{(2)}=S^TM^{(2)}S.
\end{equation}
We note that the matrix (\ref{mass eigen}) is symmetric: we can make $ M_d^{(2)}$ diagonal by choosing the \textit{mixing angle} $\theta$ properly. The expression for the angle, that we do not report here, is rather involved. The important quantities are the masses (that is, the diagonal entries of $M^{(2)}_d$), which are
\begin{equation}
    \begin{split}
        m_1^2&=\frac{1}{4}\left(7-\frac{1}{\alpha^2}\right)m^2_{\phi},\\
        m_2^2&=-m_{\chi}^2-\frac{1}{36} \left(7-\frac{1}{\alpha^2}\right)m_{\phi}^2.
    \end{split}
\end{equation}
Note that the diagonal entries (\ref{mass eigen}) are not modified by the diagonalization procedure, since $\cosh{\theta}=1+\mathcal{O}(\alpha^3)$.\\
 \indent The no-tachyon condition for the fakeon mass is $m_2^2<0$ (that is, $\alpha>1/\sqrt{7+\left(\frac{6 m_{\chi}}{m_{\phi}}\right)^2}$). Similarly, the no-tachyon condition for the physical degree of freedom is $m_1^2>0$ ($\alpha>1/\sqrt{7}$). Combining this two conditions, we get
 
 \begin{equation}\label{bound}
     \alpha>\frac{1}{\sqrt{7}},
 \end{equation}
 which is an upper bound for $\alpha$. Hence, the de Sitter expansion $\alpha\to 0$  explicitly violates the bound. In this way, the fakeon Green function satisfies the equation

 \begin{equation}
     \left(\frac{d^2}{d t^2}-m_2^2\right) G_f(t,t^{\prime})= \delta(t-t^{\prime}),
 \end{equation}
 which must be solved with the fakeon prescription \cite{classicization, ABP}

 \begin{equation}\label{green}
     G_f(t,t^{\prime})=\frac{1}{2m_2}\sin{\left(m_2|t-t^{\prime}|\right)}.
 \end{equation}
 As stated for the Starobinsky potential, the key feature of (\ref{green}) is the causality violation occurring in a typical time interval $\Delta t\sim 1/m_2$, provided that $m_2$ is real. However, the de Sitter perturbative expansion $\alpha\to 0$ violates the no-tachyon condition (\ref{bound}) and $m_2$ becomes imaginary. In other words, the violation of the bound turns the fakeon Green function into an hyperbolic sine and the violation of causality is \lq\lq propagated" up to very large time scales. For this reason, we are unable to retrieve the power spectrum of General Relativity through the decoupling limit $m_{\chi}\rightarrow\infty$: the Green function (\ref{green}) in the fundamental theory (\ref{Sinfl}) blows up so that we are unable to obtain a causal theory, General Relativity, from a strongly \textit{acausal} theory.

 \section{Conclusions}
 A theory of gravity from purely virtual particles is strongly motivated by the QFT requirements of unitarity, locality and renormalizability \cite{LWgrav}. Following the previous works in the literature \cite{ABP, CMBrunning, HighOrder, Last}, we considered the effects of said particles in the inflationary scenario by means of the cosmic RG flow approach. 
 
 We analyzed the model of quadratic inflation. Although the specific model falls in the class of potentials discussed in \cite{Last}, we have provided a detailed computation of the coefficients appearing in the tensor power spectrum. Such coefficients diverge in the limit of infinitely heavy purely virtual particles. This last feature tells us that is not possible to retrieve the well known results of General Relativity once the purely virtual particles are decoupled from the theory. The interpretation of this fact is hidden in the causal structure of the full theory, since purely virtual particles introduce (micro) causality violations \cite{Absograv, classicization, FLRW}. In particular we showed that the de Sitter limit, which is crucial to derive the perturbative expansions of the power spectra, turns the purely virtual particles into \textit{tachyons} and hence propagates the causality violations up to large time scales so that we are unable to retireve the results of the causal theory, General Relativity. We also conclude that the  model of quadratic inflation from purely virtual particles is discarded from a phenomenological point of view, likewise to what happens in the inflationary scenario driven by General Relativity. We stress, however, that these two exclusions are radically different since the former has to do with the causal structure of the theory, while the latter is related to the experimental data concerning power spectra \cite{Planck18}.

\vskip10truept \noindent {\large \textbf{Acknowledgments}}

The author warmly thanks D. Anselmi for useful discussions, P. Panci and F. Tarantelli for carefully reading the manuscript.
\vskip 1truept

\renewcommand{\thesection}{\Alph{section}} \renewcommand{\theequation}{%
\thesection.\arabic{equation}} \setcounter{section}{0}

\section{Appendix}\label{Appendix}

\label{formulas} \setcounter{equation}{0}

 \subsection{Bunch-Davies functions for the \texorpdfstring{$\phi^2$}{Lg} potential}
 
 \subsubsection{Tensor modes, \texorpdfstring{$m_{\chi}\rightarrow\infty$}{Lg}}
 The $g$ functions are
 \begin{equation}
 g_0^t=g_1^t=0,\,\,\, g_2^t=9w_0^t.
 \end{equation}
 The $w$ functions are
 \begin{equation}
     w_0^t=\frac{(\eta+i)}{\sqrt{2}\eta}e^{i\eta},\,\,\,w_1^t=0,\,\,\,w_2^t=\frac{3}{\sqrt{2}\eta}[2ie^{i\eta}+(\eta-i)e^{-i\eta}(\text{Ei}(2i\eta)-i\pi)],
 \end{equation}
  where $\text{Ei}(z)$ denotes the exponential-integral function.
  
 \subsubsection{Scalar modes, \texorpdfstring{$m_{\chi}\rightarrow\infty$}{Lg}}
 The $g$ functions are
 \begin{equation}
 g_0^s=g_1^s=0,\,\,\, g_2^s=18w_0^s.
 \end{equation}
 The $w$ functions are
 \begin{equation}
 w_0^s=w_0^t,\,\,\,w_1^s=0,\,\,\,w_2^s=\frac{3\sqrt{2}}{\eta}[2i e^{i\eta}+(\eta-i)e^{-i\eta}(\text{Ei}(2i\eta)-i\pi)].
 \end{equation}

\end{document}